## PERSPECTIVE                                                                                    Open Access

# Tilted platforms: rental housing technology and the rise of urban big data oligopolies

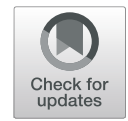

Geoff Boeing[1*] , Max Besbris[2], David Wachsmuth[3] and Jake Wegmann[4]

\* Correspondence: boeing@usc.edu
[1]University of Southern California, Los Angeles, USA
Full list of author information is available at the end of the article

**Abstract:** This article interprets emerging scholarship on rental housing platforms—particularly the most well-known and used short- and long-term rental housing platforms—and considers how the technological processes connecting both short-term and long-term rentals to the platform economy are transforming cities. It discusses potential policy approaches to more equitably distribute benefits and mitigate harms. We argue that information technology is not value-neutral. While rental housing platforms may empower data analysts and certain market participants, the same cannot be said for all users or society at large. First, user-generated online data frequently reproduce the systematic biases found in traditional sources of housing information. Evidence is growing that the information broadcasting potential of rental housing platforms may increase rather than mitigate sociospatial inequality. Second, technology platforms curate and shape information according to their creators' own financial and political interests. The question of which data—and people—are hidden or marginalized on these platforms is just as important as the question of which data are available. Finally, important differences in benefits and drawbacks exist between short-term and long-term rental housing platforms, but are underexplored in the literature: this article unpacks these differences and proposes policy recommendations.

**Policy and practice recommendations:**

- As rental housing technologies upend traditional market processes in favor of platform oligopolies, policymakers must reorient these processes toward the public good.
- Long-term and short-term rental platforms offer different market benefits and drawbacks, but the latter in particular requires proactive regulation to mitigate harm.
- At a minimum, policymakers must require that short-term rental platforms provide the information necessary for cities to enforce current, let alone new, housing regulations.
- Practitioners should be cautious inferring market conditions from rental housing platform data, due to difficult-to-measure sampling biases.

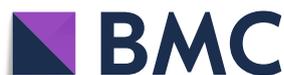

## Introduction

Information communication technologies (ICTs) shape how we live our lives and interact with one another, typically as black boxes and without explicit consent (Fields et al. 2020). The smart cities literature in critical geography has interrogated how these ICT platforms restructure capital and urban space, emphasizing the impacts of government actors contracting with corporations (Barns 2020; Caprotti and Liu 2020; Leszczynski 2016; van der Graaf and Ballon 2019). At the same time, a peer-to-peer platform urbanism has emerged in which individuals use corporations' software and algorithms to facilitate decentralized interactions. Such interactions are often unmediated by local, subnational, or national governments—but should they be?

When new technologies are introduced into society, the benefits are often obvious but the attendant drawbacks can be difficult to recognize. For example, recent consumer technologies such as smartphones and internet-connected doorbells captured the public imagination years before their privacy drawbacks became better known (Zuboff 2019). Platform urbanism is no different. Its benefits have been heralded but its drawbacks are only just starting to be identified (Koster et al. 2021; Shabrina et al. 2021). If such technologies entrench or exacerbate pre-existing inequalities, techno-utopianism is ultimately harmful and the oligopolization of user-generated "big data" by for-profit corporations becomes a matter of broad public concern.

This concern is more acute in housing than other economic sectors since, for most people, finding a permanent residence is a life-sustaining activity in a sense not true for many other consumer transactions. When a housing search fails the consequences range from bad (e.g., being unable to accept a better job in a different city) to calamitous (e.g., homelessness). In urban housing markets, technology platforms—including Airbnb and Craigslist—have created particularly strong tensions between solving and causing housing problems (Rae 2019). They enable widespread access to high volumes of short-term rental (STR) and long-term rental (LTR) housing information and should, in theory, be a boon for both homeseekers and researchers (Boeing et al. 2020b). They offer frictionless information exchange, expanded housing search radii, flexible reuse of one's home space, and a potential cornucopia of data for planners monitoring affordability. In practice, however, it is more complicated. First, user-generated data often reproduce the systematic biases found in traditional sources of housing information, intensifying the factors that fuel residential segregation (Besbris et al. 2021; Boeing 2020). Second, these platforms curate and shape information according to their owners' financial and political interests, which may not align with local public interests. Third, STR and LTR housing platforms offer different benefits and drawbacks with which policymakers have only just begun to grapple (Jiao and Bai 2019; Wachsmuth and Weisler 2018; Wegmann and Jiao 2017).

It would be one thing if housing platforms did nothing but reproduce the preexisting biases, distortions, and unequal power dynamics inherent in the underlying housing market. However they have also introduced a new concern that is relatively unfamiliar in the housing market but all too familiar in the platform economy: oligopolization. In 2021 the largest owner of multifamily rental housing in the United States owned just over 100,000 units—0.2% of the 48 million rental units overall (National Multifamily Housing Council 2021). But with housing platforms interposing themselves between brick-and-mortar housing units and many of their end users—and with the platforms subject to the same winner-take-all dynamics seen more generally with the rise of Big Tech (Lanier 2014)—the same worrying tendencies toward oligopolistic gatekeeping seen in other sectors of the economy are coming to housing.

This article reviews emerging scholarship on rental housing platforms—focusing on Craigslist and Airbnb as these are among the most researched and used LTR and STR websites—and considers how the technological processes connecting both STRs and LTRs to the platform economy are transforming cities. We discuss potential policy approaches to more equitably distribute benefits and mitigate harms of peer-to-peer platform urbanism in the rental market. ICTs are not value-neutral. User-generated information on some platforms is systematically biased against people and places of color, and for-profit corporations precipitate changes to the housing market that reduce the availability and affordability of housing in certain communities. Rental housing platforms' data may benefit certain market participants or analytics, but the evidence suggests that the benefit for some comes at the expense of society overall. ICTs alone are unlikely to mitigate existing urban inequality and policymakers must develop proactive solutions to the problems of platform urbanism.

## Reproduction of systematic biases

Information asymmetries define housing markets, and information environments shape housing search outcomes and in turn residential sorting dynamics (Akerlof and Shiller 2015; Besbris 2020). Information access is determined by social networks, life experiences, available technology, real estate agents, and demographic characteristics such as race and wealth (Besbris and Faber 2017). These combine to circumscribe homeseekers' knowledge of places and available units and exacerbate residential segregation (Krysan and Crowder 2017).

In recent years, a variety of corporations, such as RentJungle, Craigslist, and Facebook, have created centralized platforms to facilitate online information exchange in rental housing markets. Centralized exchanges *could* expose everyone to the same information, hypothetically erasing traditional information inequalities that correlate with race/ethnicity, language, age, and wealth. But, in reality, user-generated online data often reproduce the systematic biases of traditional information sources (Brannon 2017; Stephens 2013). Rental housing markets offer a clear example of this. Craigslist significantly over-represents whiter, wealthier, and better-educated census tracts (Boeing 2020). In addition, a tract's relative over- or under-representation in online rental listings is significantly associated with demographic differences in age, language, college enrollment, median rent, poverty rate, and household size (ibid.).

Beyond the volume of information, variation also exists in the type and quality of information. Craigslist rental listings in whiter or less-poor census tracts tend to include more words (i.e., information) than listings in less-white or poorer tracts (Boeing et al. 2020a). They also provide more information about available amenities, such as the presence of a washer/dryer or whether pets are allowed, that homeseekers routinely use to filter their searches. Conversely, Craigslist listings in less-white or poorer tracts have more text that focuses on prospective tenants' disqualifications—such as eviction history, criminal history, and proof of income—than listings in whiter and wealthier tracts, which tend to provide more text describing the housing unit and local neighborhood amenities (Besbris et al. 2021).

These patterns take on particular importance when considering Craigslist against the broader context of platform urbanism. Craigslist is a free, low-bandwidth web site that is easily accessible for homeseekers as well as landlords or building managers advertising available units. Yet even this platform cannot eliminate the structural factors that influence information supplies and user self-selection into certain information channels. Other platforms demonstrate similar information segregation and biases. For example, the US Department of Housing and Urban Development recently sued Facebook, claiming its rental housing platform violates fair housing law by targeting specific listings to specific users based on race, ethnicity, and religion (Porter et al. 2019). Corporate entities such as Facebook have total control over what types of information they share and where, and their platforms are the sites of clear biases that may harm consumers and benefit already advantaged groups and places. The problem is acute across the globe. For example, rental housing platforms in China frequently include inaccurate information to either disguise informal market activity or lure prospective tenants (Harten et al. 2020).

The rapid consolidation of rental housing information in the hands of a few corporations has created black boxes around information exchange, making it difficult for researchers or regulators to understand or address platform-mediated biases, steering, and sorting. These big-data oligopolies' platforms now dominate rental housing markets and the evidence suggests they amplify long-standing inequities. The ways these platforms shape market interactions and information access are inherently driven by their own corporate interests (Zuboff 2019). Planners and policymakers seeking to use them to understand housing affordability and other neighborhood processes such as gentrification will struggle to quantify and account for selection biases in community over- and under-representation. Meanwhile, homeseekers will not receive equal or holistic information about available housing, exacerbating existing sociospatial inequalities.

## The non-neutrality of housing platform data

The word "platform" conjures the image of a level surface or playing field which a disinterested party provides for others to use. In the platform economy, however, this image is misleading. Instead of providing a level playing field, technology firms curate and shape their platforms' information flows to advance their own financial and political interests. In the platform economy, data are a source of monopoly rents (Birch 2020)—a scarce factor that is economically valuable to the extent that it remains scarce.

Most consequentially, these firms exert their control over information flows to advance their financial and political interests by withholding the platform data that policymakers need to enforce current regulations, let alone craft new ones. Accordingly, a growing body of research argues that effective local regulation of STR platforms is at a minimum fraught and potentially unachievable (Ferreri and Sanyal 2018; Kerrigan and Wachsmuth 2021; Leshinsky and Schatz 2018; Wegmann and Jiao 2017; Yeon et al. 2020). For example, these scholars argue that the information necessary to regulate Airbnb hosts—minimally, their identities, the addresses of their STR listings, and the volume of activity of those listings—cannot be feasibly obtained by governments without the cooperation of Airbnb itself or private third-party firms such as Host Compliance. Airbnb remains reluctant to cooperate and has met regulation attempts with court challenges and intransigence (Holpuch 2014). At the same time, the company releases carefully curated slices of data which paint its impacts on cities and communities as favorably as possible.

Platform firms also use their control over information flows in ways that benefit the firm at the expense of users. For example, Airbnb's pricing tool recommends a nightly price to hosts. Airbnb engineers built this tool by applying "Aerosolve," the company's in-house machine learning library, to its enormous private dataset of bookings on its platform (Yee and Ifrach 2015). Airbnb does not share details of how different factors are weighted in its pricing recommendation algorithm, but industry observers widely believe that the firm systematically recommends prices lower than those that would maximize hosts' revenue (e.g. Airbnb Smart 2021; Get Paid for Your Pad 2019).

Why would Airbnb do this? As a firm, it competes with hotels and other types of tourist accommodations, and thus has an interest in keeping listing prices low, regardless of the impacts on its hosts. While individual hosts may make more money if they charge a higher nightly rate (i.e., the increased revenue per night will outweigh the decreased number of bookings which higher prices will lead to), the platform as a whole will generate more revenue if prices are lower and bookings more frequent. Airbnb wields its control over the information flowing through its platform to maximize its own utility at the expense of hosts.[1] And it does so in an opaque fashion, extracting digital rents from its control of information. This behavior is not limited to Airbnb and housing platforms. Notoriously, in April 2020, the Wall Street Journal reported that Amazon was covertly using the data it collects about third-party sales on its Amazon Marketplace to competitively guide its own product development decisions for its AmazonBasics line of consumer goods (Mattioli 2020).

Airbnb and many other housing platform firms are profit-maximizing corporations and it comes as no surprise that they flex their oligopoly power over data toward this end. The implication, though, is that housing researchers should view the data that these platforms expose to the public as a highly curated, biased fragment of a larger whole. The question of which data are made public by housing platforms must be accompanied by the trickier but equally important question of which data—and social actors—are hidden or obscured.

### Why we should be more worried about STR than LTR platforms (for now)

Platform urbanism has reshaped urban living, but nowhere are its impacts as dire as in rental housing. Housing as shelter represents a basic human need and it is therefore hard to overstate the importance of housing platforms' impacts on markets, access, and the sorting and distribution of people.

Although both LTR and STR platforms raise concerns, current research makes clear that policymakers should be more concerned, for the time being at least, by those purveying STRs rather than LTRs. Before STR platforms existed, accommodation options in cities throughout the developed world were abundant, varied widely in price and quality even in the most popular destinations, and were generally only in short supply briefly, such as during major sporting events or large conferences. If Airbnb, Vrbo, or their competitors disappeared tomorrow, out-of-town travel would continue. LTR platforms, by contrast, bring new visibility and scope—akin to what has long existed in residential sales—to the heretofore opaque rental housing market (Boeing et al. 2020b). For the time being, they do so without imposing significant costs on rental homeseekers. For instance, while Craigslist charges small fees in a small number of municipalities to landlords listing available housing, it does not profit from brokering lease agreements, and therefore does not extract excess profits from people who fulfill this basic need.

---

[1]A more benign interpretation of Airbnb's price-setting behavior is that the firm is acting in the collective interests of its hosts by resolving a collective action problem around price setting (i.e. Airbnb suggests prices which may lose hosts income in the short run but will increase the competitiveness of the STR sector relative to hotels—and thus STR host income—in the long run). But, even if this were true, Airbnb's monopolistic control of the relevant information flows, and its lack of transparency around the precise aims of its price-setting tool and other related host services, means that the firm is not acting as a neutral "platform" connecting buyers and sellers.

While LTR platforms like Craigslist reproduce systematic biases that have long existed in rental markets, STRs amplify existing inequalities and create altogether new ones. This is a consequence of how the STR and LTR rental markets interact with each other. The former exists primarily through parasitically absorbing a portion of the latter. STR industry rhetoric, as implied by the phrase "homesharing," holds that guests absorb otherwise slack residential capacity by staying in spare bedrooms of permanent housing units or occupying the entire unit while the resident is out of town. In the early days of Airbnb, this was a fair characterization of the STR market. However, robust evidence now exists that a substantial—and continually growing—portion of the STR market comprises entire units that otherwise would be used for LTR housing (Wachsmuth and Weisler 2018; Combs et al. 2020; Garcia-López et al. 2020). Housing space originally created for a city's residents is now reapportioned for use by its visitors through these platforms. This comes at a time when low-income rental housing, in the United States at least, is scarce not only within the usual list of high-priced coastal cities, but in every type of urban, suburban, and rural setting (Joint Center for Housing Studies 2020).

Although STR platforms' spokespeople claim, when arguing in the policy arena, that they act as mere facilitators of transactions between guests and hosts, they tell a different story to their users. They burnish a brand with certain connotations, such as the culturally astute Airbnb frequent traveler ("don't call her a tourist") who learns about her destination by staying and interacting with a local host. LTR platforms are different. Finding permanent housing is not something someone is eager to do frequently. A typical person seeking rental housing is looking for the right type of place in the right location at the right price, and likely cares little whether the transaction is brokered via Trulia, Craigslist, or analog means. LTR platforms therefore behave more like genuine marketplaces and less like brands.

This distinction has real consequences. Brands entail exclusivity in direct tension with transparency. In this light, it is unsurprising that Airbnb opposes fully sharing its data with researchers and governments. By contrast, Zillow readily shares its rental listing data with researchers, enabling entirely new lines of research such as questions about whether and how new market-rate multifamily buildings affect the rents of nearby existing rental housing (Asquith et al. 2019; Damiano and Frenier 2020), and whether and how STRs impact rents and housing prices (Barron et al. 2020).

Yet sharing data does not make Zillow or any other platform, like Facebook or Redfin, inherently virtuous. Zillow currently profits by selling advertising to and providing some marketing services for agents and property owners/managers, so it faces less urgency to commodify its data. But in theory, given its market power, it could create tiered access to properties such that consumers would have to pay more to see more listings. Moreover, Zillow Group, which in addition to Zillow also owns Trulia, StreetEasy, RealEstate.com, and other websites, could begin to provide other housing services (e.g., mortgage lending, direct brokerage), which would only increase the corporation's power and monopoly potential.

Like Zillow and unlike most STR platforms, LTR platforms tend not to profit from individual transactions, so the question remains how to induce STR platforms to behave more transparently. This would benefit researchers, but more urgently, it would allow cities to act on escalating calls by their residents for more robust monitoring and enforcement of new regulations to manage disruptions to quality of life and safeguard scarce rental housing.

## Toward better housing platform policy

Across both STR and LTR housing platforms, evidence of harm is growing. While research to date has overwhelmingly focused on Airbnb and Craigslist, many housing platforms operate similarly and, as a result, reproduce systematic biases and residential sorting patterns, filter and conceal information flows for their own benefit at the expense of users and communities, reapportion housing stock, and exacerbate affordability challenges and information asymmetries.

Cities can and should push back against the negative impacts of these technology firms while working with them to actually deliver the positive benefits they promise. In particular, cities should require that rental housing platforms share their listings data while protecting privacy. This is critical for understanding LTR housing affordability as the vast majority of market information now flows through online platforms. But it is also critical for STRs, where policy is more urgently needed due to their more pernicious impacts. Currently, it is nearly impossible for regulators to identify or track violations of STR regulations. Cities should require these platforms to register units, provide data on each listed unit, and report when each is rented. Some municipalities already require landlords to file leases and many counties require sellers to file deed transfers. Platforms that broker STRs should be subject to similar requirements.

Several cities offer useful models to point others toward better policy. In Los Angeles, local officials estimated that thousands of the city's Airbnb listings were illegal, but could not enforce existing regulations without better data: in response, the city recently co-developed with Airbnb a system to identify and remove illegal STR listings from the platform (Reyes 2020). The development process was contentious. City officials accused Airbnb of dragging its feet, committing insufficient resources, and unnecessarily delaying the development timeline. Airbnb in turn complained that the city failed to approve and provide the guidelines necessary to design the system in a timely manner (ibid.). Nevertheless, the system finally launched in mid-2020, allowing the city to more efficiently flag and remove non-compliant STR listings.

Likewise, Boston now requires Airbnb and similar platforms to share information monthly on listings' locations, nightly occupancy, and whether they consist of a single room or an entire unit (Enwemeka 2019). It also banned "investor units" by requiring that owners register with the city and reside in any listed STR units: Airbnb sued the city over these regulations before finally agreeing to terms (ibid.). More dramatically, Lisbon and Barcelona both took or threatened to take possession of empty short-term units and transfer them back into the long-term market at subsidized rates during the COVID-19 pandemic (Minder and Abdul 2020).

Notwithstanding these examples, cities with fewer resources may face greater challenges regulating large platforms. State or federal government intervention is likely necessary for effective policy outside of large, wealthy cities. For example, US governmental agencies could take charge of collecting, cleaning, and publicly disseminating LTR spot market data collected from Craigslist, given its value for understanding affordability conditions and the limitations of existing sources of rental market data (Boeing et al. 2020b). On the STR side, legal scholars have argued the US should modernize 1960s-era laws enacted to end racial discrimination by hotels, but that today need updating to address the well-documented discrimination against African American guests by racist hosts on Airbnb (Leong and Belzer 2017). In response to growing pressures, Airbnb has recently introduced a "City Portal" to share local indicators with cities, but specifics are scarce as the tool is brand new. Given the history of concealment and lawsuits, this likely will not solve any significant problems.

Regardless of the exact approaches ultimately taken, only with these data publicly available can policymakers begin to design platform regulations that put human lives first in the rental housing market. Cities must enforce the laws they pass and they must be willing to shut down housing platforms that refuse to comply.

#### Acknowledgements
Not applicable.

#### Authors' contributions
All authors contributed equally. The author(s) read and approved the final manuscript.

#### Funding
Not applicable.

#### Availability of data and materials
Not applicable.

#### Declaration
Competing interests.
Not applicable.

#### Author details
[1]University of Southern California, Los Angeles, USA. [2]University of Wisconsin-Madison, Madison, USA. [3]McGill University, Montreal, Canada. [4]University of Texas at Austin, Austin, USA.

## Publisher's Note

Springer Nature remains neutral with regard to jurisdictional claims in published maps and institutional affiliations.